\documentstyle[astroref-dbj,emulatedapj-dbj,psfig,epsfig]{article}
\newcommand{\COd}     {\mbox{$^{12}$CO$\;$}}
\newcommand{\COt}     {\mbox{$^{13}$CO$\;$}}

\received{** 2000} \accepted{** 2000} \journalid{337}{15 January 1989}
\articleid{11}{14} \slugcomment{}

\begin{document}

\def\th{$^{13}$}
\def\ei{$^{18}$}
\def\tw{$^{12}$}
\def\Lcs{{\hbox {$L_{\rm CS}$}}}
\def\lunits{K\thinspace \kms\thinspace pc$^2$}

\def\,{\thinspace}
\def\etal{et al.\ }


\def\kms{km\thinspace s$^{-1}$}
\def\Lsun{L$_\odot$}
\def\Msun{M$_\odot$}
\def\ms{m\thinspace s$^{-1}$}
\def\percc{cm$^{-3}$}

\font\sc=cmr10

\def\CBR{{\rm\sc CBR}}
\def\FWHM{{\rm\sc FWHM}}
\def\HI{{\hbox {H\,{\sc I}}}}
\def\HII{{\hbox {H\,{\sc II}}}}


\def\Ha{H$\alpha$}                      
\def\Htwo{{\hbox {H$_2$}}$\;$}
\def\nHtwo{n(\Htwo)}
\def\He#1{$^#1$He}                      
\def\water{H$_2$O$\;$}
\def\flecha{\rightarrow}
\def\COJ#1#2{{\hbox {CO($J\!\!=\!#1\!\rightarrow\!#2$)}}}
\def\CO#1#2{{\hbox {CO($#1\!\rightarrow\!#2$)}}}
\def\CeiO#1#2{{\hbox {C$^{18}$O($#1\!\rightarrow\!#2$)}}}
\def\CSJ#1#2{{\hbox {CS($J\!\!=\!#1\!\rightarrow\!#2$)}}}
\def\CS#1#2{{\hbox {CS($#1\!\rightarrow\!#2$)}}}
\def\HCNJ#1#2{{\hbox {HCN($J\!\!=\!#1\!\rightarrow\!#2$)}}}
\def\HCN#1#2{{\hbox {HCN($#1\!\rightarrow\!#2$)}}}
\def\HNC#1#2{{\hbox {HNC($#1\!\rightarrow\!#2$)}}}
\def\HNCO#1#2{{\hbox {HNCO($#1\!\rightarrow\!#2$)}}}
\def\HCOpJ#1#2{{\hbox {HCO$^+$($J\!\!=\!#1\!\rightarrow\!#2$)}}}
\def\HCOp#1#2{{\hbox {HCO$^+$($#1\!\rightarrow\!#2$)}}}
\def\HCOpp{{\hbox {HCO$^+$}}}
\def\J#1#2{{\hbox {$J\!\!=\!#1\rightarrow\!#2$}}}
\def\noJ#1#2{{\hbox {$#1\!\rightarrow\!#2$}}}


\def\Lco{{\hbox {$L_{\rm CO}$}}}
\def\Lhcn{{\hbox {$L_{\rm HCN}$}}}
\def\Lfir{{\hbox {$L_{\rm FIR}$}}}
\def\Ico{{\hbox {$I_{\rm CO}$}}}
\def\Sco{{\hbox {$S_{\rm CO}$}}}
\def\Ihcn{{\hbox {$I_{\rm HCN}$}}}


\def\Tastar{{\hbox {$T^*_a$}}}
\def\Tmb{{\hbox {$T_{\rm mb}$}}}
\def\Tb{{\hbox {$T_{\rm b}$}}}

\title {Chemical Evolution of the Circumstellar Envelopes of
Carbon-rich Post-AGB objects}

\author {F. Herpin\footnotemark[1]$^{,}$\footnotemark[2],
J.R. Goicoechea\footnotemark[1],
  J.R. Pardo\footnotemark[1]$^{,}$\footnotemark[3],
  J. Cernicharo\footnotemark[1]$^{,}$\footnotemark[3]}

\affil {$^{1}$Instituto de Estructura de la Materia,
   Departamento de F\'{\i}sica Molecular, CSIC, Serrano 121,
   E--28006 Madrid,  Spain}

\footnotetext[2]{Present address: Observatoire de Bordeaux, B.P. 89,
F-33270 Floirac, France, herpin@observ.u-bordeaux.fr}

\footnotetext[3]{Also visiting scientist at Division of Physics, 
Mathematics and
Astronomy, California Institute
of Technology, MS 320-47, Pasadena, CA, 91125, USA}

\footnotetext[4]
{Based on observations with ISO,
an ESA project with instruments funded by ESA Member States
(especially the PI countries: France, Germany, the Netherlands
and the United Kingdom) and with participation of ISAS and NASA.}

\begin{abstract}

We have observed with the 30-m IRAM telescope, the CSO telescope and
the ISO\footnotemark[4] satellite ({\em Infrared Space Observatory}) 
the rotational lines of
CO at millimeter, submillimeter and far infrared wavelengths in the direction
of C-rich stellar objects at different
stages of evolution :  CRL 2688 (a very young Proto-Planetary Nebula),
CRL 618 (a Proto-Planetary Nebula), and NGC 7027 (a young Planetary Nebula).
Several changes in the longwave emission of CO and other molecules
are discussed here in relation
with the degree of evolution of the objects. In the early stages,
represented by CRL 2688,
the longwave emission is dominated by CO lines.
In the intermediate stage, CRL 618, very fast outflows are present which,
together with the strong UV field from the central star, dissociate CO.
The released atomic oxygen is seen via its atomic lines, and allows the
formation of new O-bearing species, such as \water and OH. The
abundance of HNC is enhanced with respect to HCN as a result
of the chemical processes occurring in the
photo-dissociation region (PDR). At this stage, CO lines and [O$\small I$]
lines are the dominant coolants, while the cooling effect of [C$\small{II}$]
is rising. At the Planetary Nebula stage, NGC 7027, large parts of the
{\em old} CO AGB material have been reprocessed. The spectrum is 
then dominated by atomic and
ionic lines. New species such as CH$^{+}$ appear. Water has probably 
been reprocessed in OH.


\end{abstract}

\section{Introduction}

Solar type stars remain in the asymptotic giant branch (AGB) for about
$\sim 10^{6}-10^{7}$ years, losing mass through stellar winds ($\dot{M}
\sim 10^{-4}-10^{-7}$ M$_{\odot}/$yr, ejection velocity $\sim 5-25$ \kms,
see, e.g., Loup \etal 1993). During this phase the ejected material
progressively forms an expanding circumstellar envelope (CSE).
Molecular species are easily formed in the innermost
regions of the CSE. Some of these molecules aggregate onto  dust
grains which under the action of radiation pressure accelerate and push
the remaining gas producing an expanding dusty molecular envelope.
UV photons from the interstellar radiation field dissociate
gas-phase molecules in the external layers of the envelopes
  allowing new chemical reactions and the production of
new molecular species. These chemical changes have been very well studied
through observations and models in the case of IRC+10216, the prototypical
AGB carbon-rich star (Glassgold 1996; Cernicharo, Gu\'elin \& Kahane 2000
and references therein).
Additional changes in the composition of the CSE
do occur when the central object starts its evolution towards the white dwarf
stage. A large UV field arises from the central (much hotter) star
and at the same time, high velocity winds
appear and interact with the AGB remnant (Cernicharo et al. 1989; Kwok
2000). Objects in this phase are called proto-planetary nebulae
(hereafter {\em PPNe}).
During this period (1000$-$2000 years; Bujarrabal et al. 2001), the 
almost spherical
symmetry typical of  the AGB stage disappears and is replaced by a
complicated geometrical structure (most PPNe are elliptical, bipolar or
quadrupolar, Zuckerman \& Aller 1986; Frank \etal 1993).
The chemical composition is also strongly affected by these processes:
O-bearing molecules can be formed in C-rich objects
(Herpin \& Cernicharo 2000), and complex organic molecules, which are not
observed in IRC+10216, are efficiently produced (Cernicharo \etal 2001 a\&b).

In order to better understand this evolution,
we present in this paper a comparative study of the millimeter, submillimeter
and far-IR CO line emission from 3 objects representing different stages
of this fast transition: CRL 2688, a very young PPN, CRL 618, a PPN, and
NGC 7027, a young planetary nebula (PN). The different excitation coSnditions
of the observed CO
lines allow to probe different layers in the CSEs. In particular,
we study the CSE remnant wind and the higher velocity winds.
The CSE component is the "normal" AGB wind, and the wind component refers
  to a higher velocity wind which is probably not present during the
AGB phase and which contributes to changing the morphology from 
spherical to bi-polar or other non-spherical geometries. The observations are
presented in section 2. The representativity of each selected source is
discussed in section 3. Section 4 is devoted to the analysis of the
spectra of IRC+10216 as a reference of the AGB phase. In section 5
we discuss the wind structure of the young PPN CRL 2688. Section 6 is
devoted to CRL 618 and section 7 to the analysis of NGC 7027 observations.
An overall comparative discussion of the four objects is then presented in
section 8.

\section{Observations and Data Analysis}

The IRAM-30m observations of the CO 1$-$0 and 2$-$1 lines were performed in
September 2000 using four SIS receivers with high ($>$0.90) image sideband
rejection.
The pointing and focus
were checked using either Saturn or Jupiter, using the continuum
emission of the sources and was found to be accurate to within 
2$^{\prime\prime}$.
The typical system temperatures during these observations were 250 K
(CO 1$-$0) and 400 K (CO 2$-$1).
The observations used the wobbler system which provides
very flat baselines. The spectrometers were a filterbank (at 115 GHz)
and an autocorrelator (at 230 GHz) set to cover a band of 512 MHz with
frequency resolutions of 1 and 1.25 MHz respectively (velocity resolutions of
2.60 and 1.63 \kms $\;$ respectively).

The CO 3$-$2, 4$-$3, 6$-$5 and 7$-$6 observations were performed with 
the 10.4 m
telescope of the Caltech Submillimeter Observatory at the summit of Mauna
Kea (Hawaii) in September and October 2000 using three different receivers
based on helium-cooled SIS mixers operating in double-sideband mode (DSB).
The typical system temperatures during these observations were 750 K (CO 3$-$2)
and 5000$-$6000 K (CO 4$-$3, 6$-$5 and 7$-$6). For further details about the
receivers see Kooi \etal (2000) and references therein. The backends 
were two different
acousto-optic spectrometers. One of them covers
a bandwidth of 500 MHz with 1024 channels, and the other one covers 
1.5 GHz with
2048 channels thus providing respectively the following velocity resolutions
(in \kms): $146.38/\nu$ and $219.57/\nu$, with $\nu$ in GHz. The
instantaneous band of the receivers, however, is $\sim 0.9$ GHz, thus
limiting the validity of the data from the 1.5 GHz AOS to that range.

Since some of the CSO observations were performed shortly after a new 790$-$920
GHz receiver
was installed for the first time at the telescope, its pointing and focus
were carefully determined during previous runs earlier in the year. Then,
the pointing of this and the other receivers was checked during our 
observations
using Jupiter and Saturn first and then the own CO emission of each object.
Taking into account that this emission is somewhat extended, the pointing is
considered accurate to within 4$^{\prime\prime}$ (CO 3$-$2, 4$-$3 
lines) and 2$^{\prime\prime}$ (CO 6$-$5 and 7$-$6
lines).
The observations were also performed by wobbling the secondary reflector by
60$^{\prime\prime}$. Proper corrections were finally
applied  to the data to compensate for the use of only one load (ambient
temperature) in the
chopper-wheel standard calibration of the CSO that relies upon the
Penzias-Burrus (1973) approximation.

For the comparative discussion that we make further in this paper,
fits of the CO mm/submm observations (shown
in Fig. 1) of the three objects have been performed to derive the
flow velocity of the different components and the integrated
intensity ratio $R_{I}$
between the emission of the CSE remnant and that of the wind for each 
transition.
As angular resolution changes when observing at different frequencies
and using different telescopes, the relative contribution to the bulk of the
detected emission from the CSE remnant (extended) and the high velocity
winds (unresolved) will be different. Hence, $R_I$ outlines the wind
excitation of the  different flows across the
envelope (cf. Tables 1,2 and 3).

The ISO LWS Fabry-Perot data
($\lambda / \Delta \lambda \sim 9500$) of CRL 2688
were obtained during orbits 366 and 369. The
integration times are 3400 and 5800 sec. The spectral
resolution is 0.015 $\mu$m. The LWS grating
mode (43$-$196.7 $\mu$m, $\lambda / \Delta \lambda \sim 200$) has a spectral
resolution of 0.29 $\mu$m for the 43$-$93 $\mu$m range and 0.6 $\mu$m for the
80$-$196 $\mu$m range. LWS01 grating spectra for CRL 2688 and CRL618
were taken during orbits
21 and 688 respectively (total on-source times of 1218 sec and 5360 sec,
respectively).
The J=18$-$17 line of HCN was observed towards CRL 2688 during orbit 154
with the LWS grating mode and an integration time of 914 sec (see Fig. 4).
Finally,
the LWS grating spectra of NGC7027 taken by ISO during orbits
21, 342, 349, 356, 363, 377, 537, 552, 559, 566, 579, 587, 594,
601, 706, 713, 720, 727, 734, 741, 755, 762, 769, 776 and 783 were averaged.
The total on-source time was 53409 sec. All data were processed following
pipeline number 9 and treated with
ISAP\footnotemark[3] to remove glitches and fringes.

\footnotetext[3]{The ISO Spectral Analysis Package (ISAP) is a joint 
development
by the LWS and SWS Instrument Teams and Data Centers.
Contributing institutes are CESR, IAS, IPAC, MPE, RAL and SRON.}

\begin{figure*} [p] 
  \begin{center} 
     \epsfxsize=14.cm 
     \epsfbox{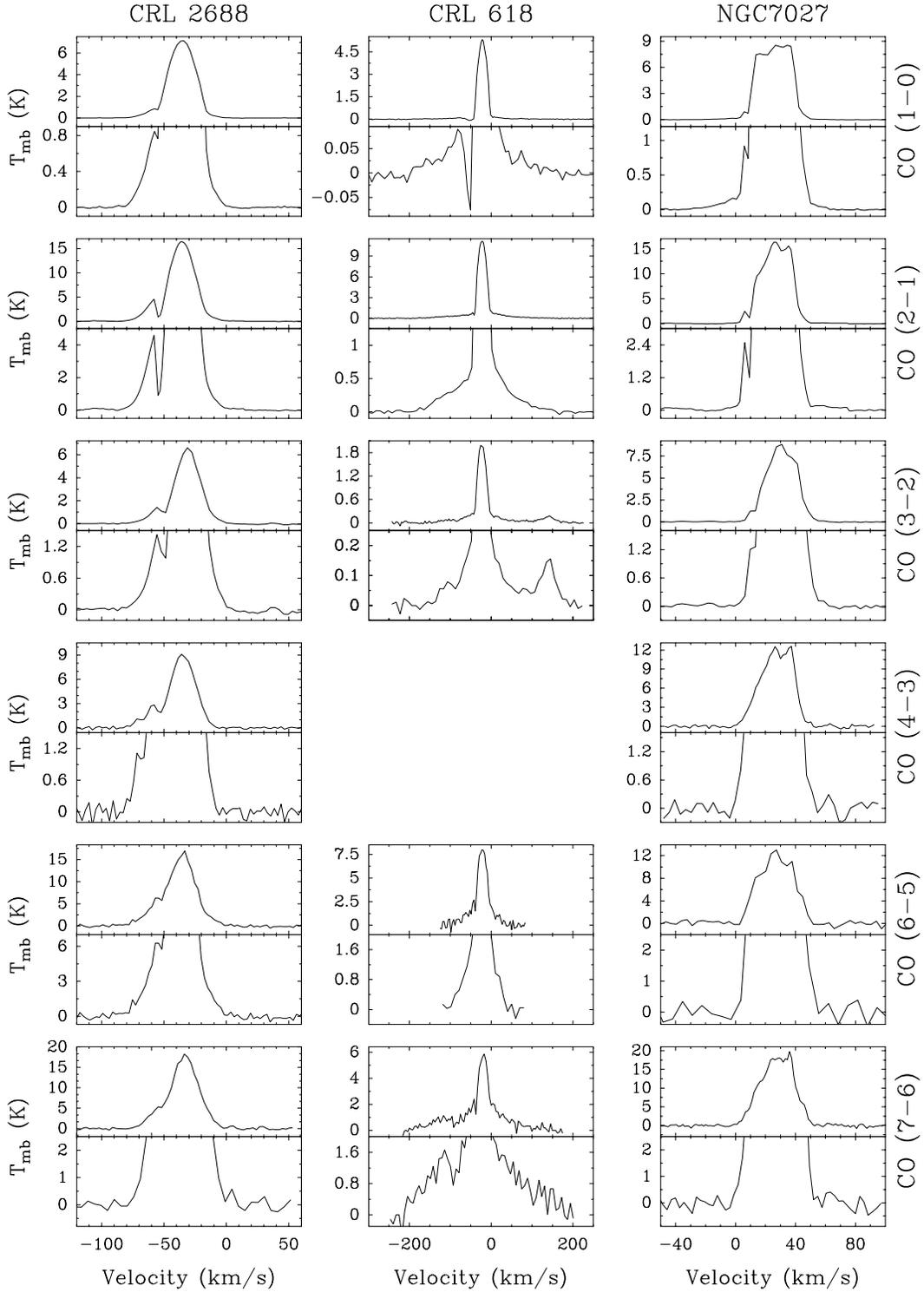} 
  \end{center} 
  \caption []{\footnotesize{Spectra for the \COd J=1$-$0, 2$-$1, 3$-$2, 
4$-$3, 6$-$5 and 7$-$6 transitions
for CRL 2688, CRL 618 (no J=4$-$3 observation)
and NGC 7027. The main beam temperatures are in K and
the velocity in \kms. Baselines were applied. The 1$-$0 and 2$-$1 lines
are IRAM
30m observations; the others were made with the CSO; lines are smoothed.
For each line, the upper part of the caption shows the whole profile, and
the lower part shows the high velocity emission with an expanded vertical
scale. Note that the emission present on the red part (at 141 \kms) of the
CO (3$-$2) emission from CRL 618  is the HC$_{3}$N (J=38$-$37) line emission
(at 345.609 GHz). The conversion factors Jy/K are respectively 4.3, 
4.2 and 40 for the 1$-$0, 2$-$1 observations and rest of 
observations.}} 
  \label{lw} 
\end{figure*}

\section{Representativity of the source sample}

 In the next sections we try to identify and discuss the changes that ocurr through
the transition of a C--rich  AGB star to the PN stage. The aim of this work is to 
trace the evolution of the circumstellar chemistry in the late-AGB and
post-AGB phases.
This study is based on the analysis of the spectral differences observed
in 4 objects: IRC+10216, CRL 2688, CRL 618 and NGC 7027.
We thus have to relay upon the assumption that  IRC+10216 will evolve into
an object  similar to the Egg Nebula, and then to an object  similar to
CRL 618 which will finally evolve to a planetary nebula that looks like
NGC 7027.
The ideal, but unfeasible approach, would be to observe a single 
object evolving through the post-AGB phase. 
In addition, our sample objects should be ``typical''
in the sense that each one should  represents the averaged characteristics
that the bulk of objects in the same stage of evolution.

However, CRL 2688, CRL 618 and NGC 7027 seem to be quite unusual
objects in some properties, extreme cases among all carbon-rich AGB and post-AGB objects.
All are very bright in molecular lines, but while IRC+10216 is a relatively
close source (110-150 pc, Groenewegen \etal 1998, Crosas \& Menten 1997),
the other three objects are  5-10 times further away (Acker \etal 1992,
Bujarrabal \etal 1994, Sahai \etal 1998). This may twist
the comparison between IRC+10216 and the other sources, as distance influences line 
fluxes and profiles.

Recent works estimate a  $\leq 2$ M$_{\odot}$ progenitor mass  for IRC+10216
and place  its core mass to 0.6 M$_{\odot}$ (Kahane \etal 2000).
Theoretical calculations suggest a progenitor main sequence mass of $\sim 3$ M$_{\odot}$
for both CRL 2688 and CRL 618 (Speck, Meixner and Knapp 2000), and 3-4
M$_{\odot}$ for NGC 7027 (Sahai \etal 2001).
IRC+10216 therefore, seems likely to evolve to the PN phase rather more 
slowly than the others, and this may affect the chemistry of the envelope. 
Anyway, several parallelisms exist between these objects: dense, optically 
thick dusty circumstellar envelopes and shell-like density enhancements
observed as incomplete arcs (Men'shchikov \etal 2001, Mauron \& Huggins 1999). 
These circumstellar concentric arcs, extremely pronounced in IRC+10216,
are also detected in NGC 7027 (see Kwok, Su \& Stoesz 2001).
More generally, it has been proposed  that further evolution of
CRL 2688 will lead to morphologies similar to that observed in NGC 7027
(specially the ionized nebula and H$_2$ region; Cox \etal 1997).

The Egg Nebula is quite anomalous compared to more usual carbon-rich post-AGB
objects. Its infrared spectrum is very different from that of any other 
carbon-rich post-AGB object, since it appears to be optically thick even at
30 $\mu$m (Omont \etal 1995). 
More common are the most extreme dusty carbon-stars such as the small group
of sources discussed  by Volk \etal (2000).

CRL 618 is quite similar to CRL 2688 in that both have massive molecular
envelopes  and central bipolar outflows (Meixner, Fong \& Sutton 2001). 
However, CRL 618 shows an unusual rich circumstellar chemistry that
could be related to its advanced stage of evolution (Herpin \& Cernicharo 2000).

NGC 7027 is quite unusual among planetary nebulae in that it has a relatively
massive circumstellar envelope beyong the ionized region. Its PDR is one
of the densest and warmest in the  PN sample
studied by Liu \etal (2001). In fact, the ionized
region is known to be relatively low in mass ($\leq 0.2$ solar masses, smaller
than the typical values of 0.3-0.5 M$_{\odot}$ of the ionized region of PN)
while the surrounding mostly molecular gas totals 2-4 solar
masses.

In conclusion, the objects selected here are rather peculiar
in some properties that may influence their post--AGB evolution. However, 
a comparative study of their molecular envelopes could be useful
to identify the distinct chemistries induced by the general physical
changes produced by the different degrees of evolution, and this is 
the aim of this paper.

%
%
\begin{figure*} [ht] 
  \begin{center} 
     \epsfxsize=15cm 
     \epsfbox{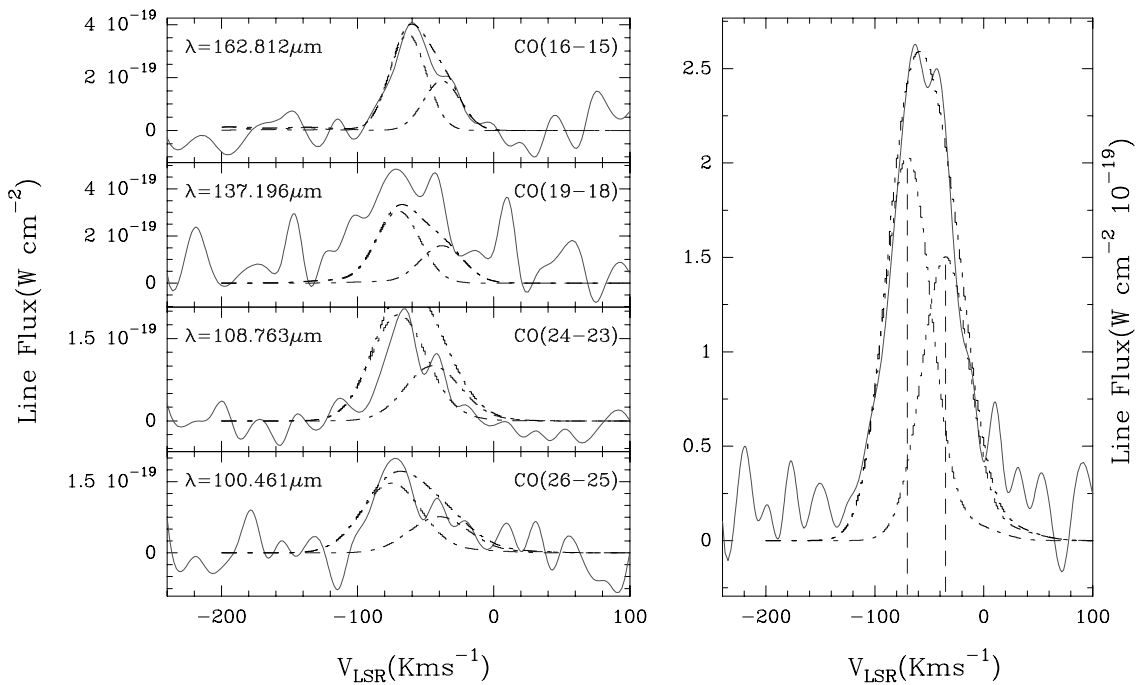} 
  \end{center} 
  \caption []{LWS/FP observations of \COd lines in CRL 2688 (left panel) and
average spectra of all observed lines (right panel).
The intensity scale corresponds to the
continuum normalized flux rescaled to the values obtained with the LWS
gratings for the same lines. The wavelength of each transition is indicated
at the top left corner of each panel. The right panel shows the average
spectra of the CO lines. Dashed lines show the fit produced by our LVG model as
described in the text (with a center velocity precision of $\pm 15$ \kms).} 
  \label{lw} 
\end{figure*}

\section{IRC+10216 : the Prototypical AGB Carbon-Rich Object}

To introduce this study, it is useful to present the main features
of a typical object at the end of the AGB stage.
IRC+10216 is the brightest AGB carbon rich star. Its estimated distance is
110$-$150 pc (Groenewegen 1997, Crosas \& Menten 1997).
It is believed to be in the final stage of red giant evolution and
exhibits a bipolar reflection nebula (Skinner, Meixner \& Bobrowsky 1998).
The temperature of the central object
is 2000 K, and the expansion velocity of the CSE is 14.5 \kms $\;$
(Skinner, Meixner \& Bobrowsky 1998; Cernicharo et al. 2000).
IRC+10216 has an extended CSE where more than 50 molecular species have
been detected (see line surveys by Cernicharo \etal 1996,
Cernicharo et al. 2000).
This object has a rich carbon chemistry and many of the
species detected in its molecular envelope are carbon chain radicals which
are formed in the external layers of the CSE (Cernicharo \etal 1996;
Cernicharo et al. 2000). The
innermost regions of the envelope are dominated by a chemistry at
thermodynamical equilibrium (Tsuji 1973), where CO, HCN, C$_{2}$H$_{2}$,
SiO are formed. As shown by Cernicharo \etal (1996), the
far-IR spectrum consists of strong dust emission and several molecular
emission lines from \COd (J=14$-$13 to J=39$-$38),
\COt ([\COd$/$\COt]$\sim 45$), HCN (J=18$-$17 to J=48$-$47), H$^{13}$CN
and vibrationally excited HCN. The
HCN lines have intensities similar to those of CO,
and [HCN$/$CO]$=1/10$, proving that HCN is the
main coolant in the innermost region of the CSE. Only one ion,
HCO$^{+}$, has been detected (Lucas \&
Gu\'elin 1999) and there is no [C$\small{II}$]  line emission.
The mid- and near-infrared spectra are dominated by the ro-vibrational
lines of C$_2$H$_2$ and HCN (Cernicharo et al., 1999).
The most abundant molecules through the CSE are CO, HCN and
C$_{2}$H$_{2}$ (Fuente, Cernicharo \& Omont 1998,
Cernicharo et al. 2000).

The molecular observations of this object
are well explained by a photochemical model (Glassgold 1996)
considering that, due to the low effective temperature
of the central star, the only UV photons
available in the outer CSE come from the interstellar radiation field.

\section{A young PPN: CRL 2688}

CRL 2688 (the {\em Egg Nebula}) is a very young PPN (it has left the
AGB phase probably only about 100 years ago) with an effective
temperature around 6600 K (spectral type F5, Justtanont \etal 1997)
and luminosity close to $1.8 ~10^{4}$ L$_{\odot}$ (Jura \& Kroto 1990).
The non-detection of [O$\small I$] and [C$\small{II}$] (Cox \etal 1996)
indicates that radiation from the central star has not
yet formed a PDR. The molecular gas is in expanding fragmented shell
structures, and shocks are thought to heat it (Cox \etal 1996). The
AGB CSE expansion velocity is 20 \kms. CO emission has been seen as far as
42$^{\prime\prime}$ from the central object. The weaker \COt emission 
extends over 20$^{\prime\prime}$
(Yamamura \etal 1996).
Sahai \etal (1998) detected a moderate velocity wind (hereafter {\em MVW})
of 40 \kms $\;$ ~with an angular extension of 10$^{\prime\prime}$.
High velocity gas emerging in
two different directions from the central
star has been detected by Cox \etal (2000)
in the CO J=2$-$1 emission line.
The size of the emitting region is 4$^{\prime\prime}$ with
extensions in both the north-south and east-west directions.
Cox \etal (2000) have shown that this high velocity gas is closely
related to the \Htwo emission at 2 $\mu m$.

%
\begin{figure*} [p] 
  \begin{center} 
     \epsfxsize=16cm 
     \epsfbox{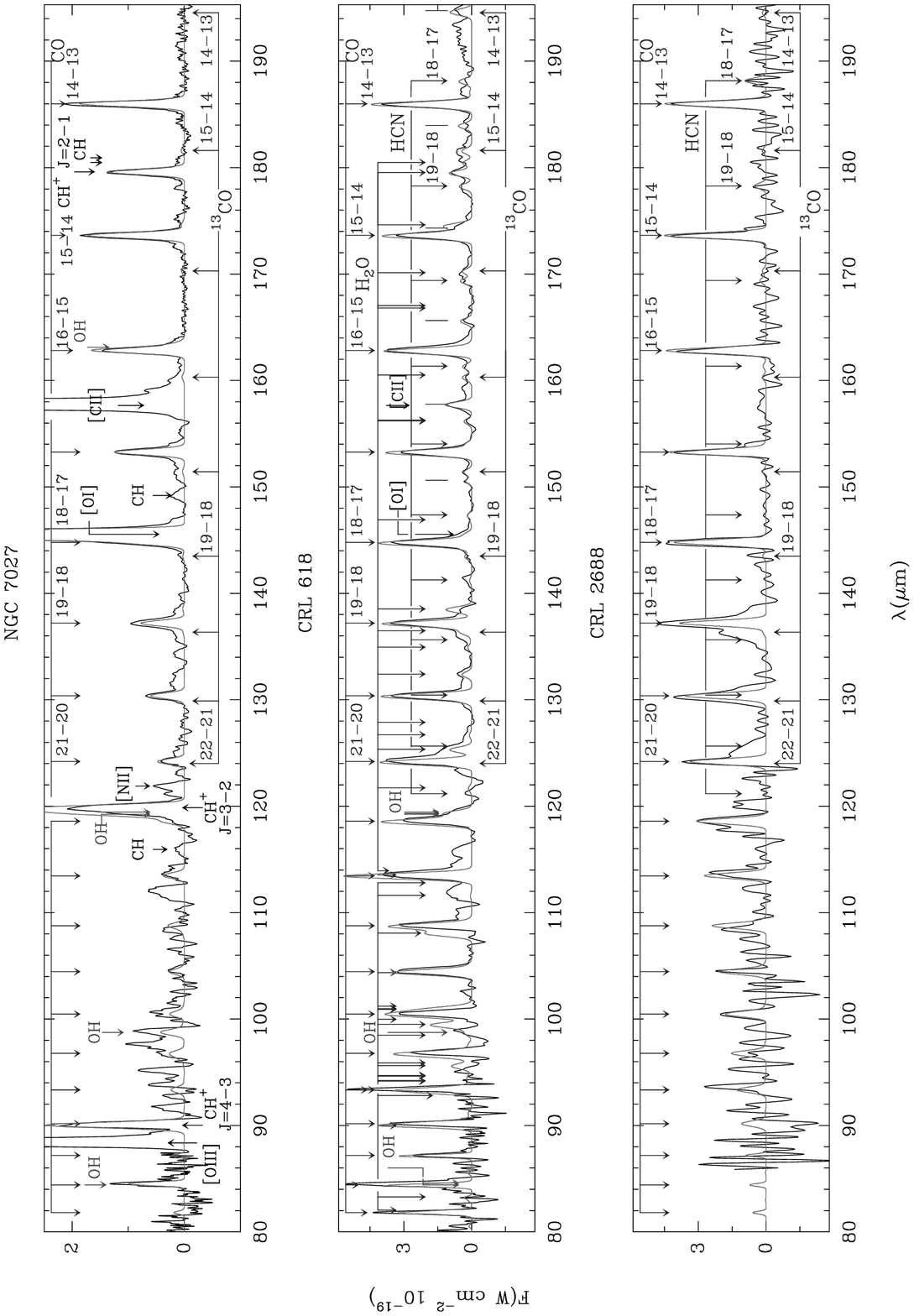} 
  \end{center} 
  \caption []{\footnotesize{Continuum subtracted LWS spectra of CRL 2688 
(bottom caption),
CRL 618 (middle caption) and NGC 7027 (top caption). The result of our models
is shown by the continuous grey line.  The lines of \COd, \COt, HCN, \water,
CH$^{+}$ and OH are indicated by arrows while those of HNC in CRL 618
are indicated by vertical lines (from J=22$-$21 at 150.627 $\mu$m to 
J=17$-$16 at
194.759 $\mu$m). The [C$\small{II}$] and [O$\small{III}$] lines
are not included in our
models; the plots indicate gaussian fits to these features.}} 
  \label{lw} 
\end{figure*} 
%
%

\subsection{Low-excitation CO lines}

The profiles of all low excitation CO lines (see Fig.1)
are characteristic of optically
thick emission. However, while the J=1$-$0, 2$-$1, and 3$-$2 lines
probably trace the
extended AGB remnant envelope of CRL 2688, higher J transitions are more
sensitive to hotter gas from the inner layers of the envelope and
hence of lower spatial extent. As J increases, the line opacities
in the AGB CSE decrease and the emerging profiles change accordingly.

 From the CO observations of CRL 2688 and the line parameters given
in Table 1 we can distinguish two main outflows. The main one,
the AGB CSE remnant, is centered at $-$35.4 \kms $\;$ and covers around
20 \kms $\;$ (in good agreement with Young \etal 1992).
The second outflow has a moderate velocity extent, $\simeq$50 \kms, and
is centered at $-$60 \kms. The intensity of the red wing of
this MVW is weaker than its blue counterpart in the CO
lines. We believe that this behavior is
produced by AGB CSE remnant absorption of the MVW red wing.

An absorption dip is detected in the blue part of the mm and submm CO lines
at $\simeq$$-$55 \kms, just at the terminal velocity of the AGB CSE.
It suggests that the low-velocity wind absorbs radiation from the 
inner and faster wind.
Young et al. (1992) found that this absorption has a velocity width of
$\simeq$1 \kms $\;$ in the low-J lines. We found the same
value for the J=6$-$5 and 7$-$6 CO lines. This would
suggest that the microturbulence through the AGB CSE is
nearly constant and equal to 1 \kms.

The ratio $R_{I}$ goes from 0.08 (J=1$-$0) to 0.24 (J=7$-$6),
stressing the importance of the MVW emission in the intermediate-J CO lines.
The dilution of the MVW emitting region in the telescope beam will change
with J. In addition, line opacities for the AGB CSE remnant and the MVW
will also change differently with J. Hence, above a given rotational
transition, the emission from the AGB CSE remnant will be
weaker than that from the MVW. Observations of high excitation lines
are thus needed in order to follow up the CO emission from the inner envelope.

\begin{table*} [h]
   \caption{ \label{tablex}Table of the parameters for the different flows in
     CRL 2688. The uncertainty on $v_{main}$
     and $v_{wing}$ is 0.3 \kms.
     $R_{I}$ is the wing and main integrated intensities ratio.}
{\small {\begin{tabular}{l|c|c|c|c|c|c}
  & \multicolumn{6}{c}{\bf{CRL 2688}} \\
  {\emph CO parameters} & 1$-$0 & 2$-$1 & 3$-$2 & 4$-$3 & 6$-$5 & 
7$-$6  \\ \tableline
$v_{main}$ (\kms) & 19.3 & 20.0 & 23.9 & 24.1 & 23.4 & 23.3 \\
$v_{wing}$ (\kms) & 50.0 & 52.0 & 50.7 & 40.3 & 41.6 & 33.3 \\
$ R_{I}$ & 0.08 & 0.08 & 0.11 & 0.16 & 0.24 & 0.24 \\
  $\Delta$$R_{I}$ & (0.01) & (0.01) & (0.01) & (0.02) & (0.02) & 
(0.03) \\ \tableline
\end{tabular}}}
\end{table*}

\subsection{Far-IR emission}

Far-IR observations reveal the inner and hotter regions of the CSE
where important chemical and physical processes are changing the
structure of the AGB CSE. Rotational lines of \COd (J=14$-$13 to 26$-$25)
are clearly detected while the emission from \COt is very weak.
HCN is clearly detected in the AOT LWS02 grating spectrum of its J=18$-$17
transition (see Fig. 4, line detected with $S/N\sim4$; all the 
sub-peaks are smoothed noise). Note that some CO lines present broad
wings  at the grating resolution (see J=19$-$18, J=20$-$19, J=21$-$20) that are
probably instrumental artefacts resulting from the defringing of the LWS
data.

Several high-J CO (from J=16$-$15 to 26$-$25) lines were observed in 
Fabry-Perot
mode with the LWS (see Fig. 2 for the J=16$-$15, J=19$-$18, J=24$-$23 
and J=26$-$25
lines). The fluxes were divided by the continuum and rescaled to the values
obtained for the same lines with the LWS grating. To derive more valuable
information about kinematics and because the excitation and opacity is rather
similar in all these lines, we computed an average spectrum (better S/N ratio)
of all the individual lines (see right panel in Fig. 2).
Study of this average CO LWS/FP spectrum reveals a much broader line than
what we could expect from the AGB CSE remnant and  centered at
$-$65 \kms. It could prompt that all the emission is coming from the MVW.
In order to estimate the CO emission at high-J from the AGB CSE remnant and
from the MVW, we have fitted two gaussians to the average spectrum of
Figure 2. We obtain velocities of $-$70$\pm$10 \kms $\;$ and 
$-$35$\pm$10 \kms $\;$
(see on the same figure the modelled emission from our
{\em Large Velocity Gradient} (LVG) calculations -discussed below). These
velocities agree, within the FP absolute wavelength calibration,
with the MVW and AGB CSE velocities respectively. Obviously, higher spectral
resolution observations are required to confirm both components
(i.e., HERSCHEL observations with HIFI).
In the following we assume that both components
emit in the high-J lines of CO and we will try to derive the relative
contribution of each one.

Obviously, the average spectrum does not directly provide information on
the column density of CO. In order to derive this parameter we have fitted
the CO J=16$-$15 emission observed with the FP (the line with the best
S/N ratio) with the two velocity components derived from the average
spectrum. We
then extended the result to the other lines (Fig. 2, left panel).
Globally, the lines are well fitted by our model, with column densities
of 85\% (feature at $-$70 \kms) and 45\% (feature at $-$35 \kms) with 
respect to
the values retrieved from fitting the grating spectrum.

The red counterpart of the emission at $-$70 \kms $\;$ is
not detected. Our lines suggest that one part of the emission
(at $-$70 \kms) comes from a
wind with an outflow velocity of about 35 \kms,
probably the MVW of Sahai \etal (1998).
As the $-$70 \kms $\;$ emission comes from a region where large shocks occur
(interaction with the AGB remnant envelope), and the column density is
thought to be larger, this emission appears stronger.

The other component (at the stellar velocity)
is consistent with the wind velocity derived from our CO mm and submm
observations (see Table 1). The widths of the emissions in each CO line show
that the velocity dispersion is probably important in the excitation region.

In order to model the observed emission we have used an LVG code which
divides the emitting source into several components with
different solid angles (defined by $\theta$). For each component
we have fitted the column density of the molecule under consideration, the
H$_2$ volume density and the temperature. Hence, the molecular
abundance will depend on the layer thickness  which cannot be derived
in this retrieval scheme. While these assumptions could be correct  for
the high velocity wings and the PDR (CRL 618, NGC 7027), the component
corresponding to the AGB remnant is poorly approximated as the whole
CSE is reduced to a layer with a given average column density,
temperature and density. This AGB remnants has been modelled in detail
by Bujarrabal \& Alcolea (1991).

The same basic physical
parameters have been applied to \COt and HCN and we then
only varied the column
densities (see parameters in Table 4). For all the objects,
  model components were selected according to previous observational
  and theoretical works and our millimeter and submillimeter observations.
  The particular temperatures,
solid angles, etc. are also derived from these studies, even if in
some cases we adapt the parameters to correctly fit the data. Of course,
column densities for each layer are sensitive to the excitation temperature.
 
According to other authors and our study of the low-J CO
line emission, the adopted geometry model and physical parameters of
each component are :

(i) An inner region representing the fast wind from the star
(Sahai \etal 1998) and modelled as
a layer of $\theta$= 0.3$^{\prime\prime}$ , expansion velocity
of 100$-$200 \kms,
temperature of 800 K, and a density of $5 ~10^{7}$ cm$^{-3}$ .
The \COd column density ($N$) is $2 ~10^{18}$
cm$^{-2}$ and
the \COt and HCN abundances relative to \COd are $<1/25$ and $<1/100$
respectively. This hot component is needed to fit correctly
the high-J CO lines, as the single 400 K component of Cox \etal (1996)
does not seem to be sufficient in our view.

(ii) As this fast wind slows down when shocking with the
outer parts of the envelope, we have considered intermediate layers of
larger sizes. Parameters are from Cox et al. (1996, 2000):
a layer of 500 K, $\theta$=2$^{\prime\prime}$ , 100 \kms $\;$ 
expansion velocity, and density
of $5 ~10^{7}$ cm$^{-3}$. The \COd column density has decreased to
$7 ~10^{17}$ cm$^{-2}$ and the HCN$/$CO abundance ratio is $<1/30$.
An additional component is required where the wind has slowed down to 60 \kms,
with a temperature of 400 K, $\theta$=3$^{\prime\prime}$, density of 
$5 ~10^{7}$
cm$^{-3}$ and $N$(\COd)$=5 ~10^{17}$ cm$^{-2}$. \COt and HCN have the
same relative abundance ratios than in the previous layer.

(iii) A moderate velocity wind region (Sahai \etal 1998) with 
$\theta$=15$^{\prime\prime}$, temperature of 250 K,
density of $5 ~10^{7}$ cm$^{-3}$, velocity of 40 \kms, and \COd column
density of $4.5 ~10^{16}$ cm$^{-2}$.
The relative abundances for the other species are the same as for previous
layers.

(iv) The AGB remnant CSE where low-excitation CO emission extends 
beyond 40$^{\prime\prime}$ (Cox \etal 1996, Yamamura \etal 1996)
and that we have modelled as a layer of  25$^{\prime\prime}$ at 50 K,
expansion  velocity of 20 \kms, $N$(\COd)$=3.5 ~10^{16}$ cm$^{-2}$.

The global fit is shown in Figure 3 and reproduces reasonably well the
LWS-ISO grating observations of CRL 2688. We also applied the same 
model to the mm and submm CO lines. Components (iii) and (iv) 
mainly fit the low-J CO lines (millimeter and submillimeter lines). 
Observed and predicted intensities are in good
agreement (see Table 7).

\subsection{Changes since the previous stage}

With respect to the previous AGB stage, represented by
IRC+10216, we see that the molecules emitting in the far-IR
are basically the same, but with different relative intensities.
This is probably due to the fact that the innermost region of the AGB
remnant envelope has been already pushed away in these initial stages of
the evolution towards the PPN phase. Hence, all the ro-vibrational
lines of HCN have disappeared, not necessarily due to a depletion of
this molecule but to the lack of sufficiently high temperatures and densities.
This fact, together with the lack of [O$\small I$] and [C$\small{II}$], makes
CO the main coolant of the envelope. The observational evidence
of winds faster than those in the previous stage indicates that the gas in the
inner layers is now heated by shocks, where the high-J CO lines excitation
is favored. The presence of polyynes, as shown by the ISO mid-IR
spectrum of this object (Cernicharo et al., 2001a), indicates that the
envelope has also started an important chemical evolution when compared to
IRC+10216. The high velocity winds and the increasing UV field from the
central object are at the origin of this new chemistry, and this will
be enhanced in the next stage (CRL 618).

\section{A PPN object: CRL 618}

CRL 618 is one of the few clear examples of an AGB star in the transition
phase to the Planetary Nebula stage. There is evidence that this
object must be a $\sim 200$ years old PPN (Bujarrabal \etal 1988).
It has a compact HII region created by a hot central star
(30 000 K, spectral type B0, Justtanont \etal 1997).
CRL 618 is seen as a bipolar
nebula at optical, radio and infrared wavelengths (Carsenty \&
Solf ~1982, Bujarrabal \etal ~1988, Cernicharo \etal 1989,
Neri \etal 1992, and Hora \etal ~1996).
The expansion velocity of the outer envelope
is around 20 \kms. CO observations by Cernicharo \etal (1989)
evidenced the presence of a high-velocity
outflow ($\sim$200 \kms). The UV photons from the central star and
this high velocity wind perturb the circumstellar envelope (CSE),
producing shocks and photodissociation regions
  which modify the physical and chemical conditions of the gas
(Herpin \& Cernicharo 2000; Cernicharo \etal 2001a\&b).

The molecular content of CRL 618 (Cernicharo et al. 2002, in preparation) is
characterized by a large abundance of cyanopolyynes, acetylenic 
chains, methane,
methylpolyynes and benzene (Cernicharo et al. 2001a\&b): this PPN
is a very eficient organic chemistry factory. The detection of long carbon
chains in both CRL 618 and CRL 2688 suggests that the input of a UV radiation
field from the evolving central object or/and the action of high velocity
winds on the slowly expanding AGB remnant modify the CSE chemistry allowing
the formation of new and more complex molecules.

\smallskip 
\vspace{7.5cm} 
\includegraphics{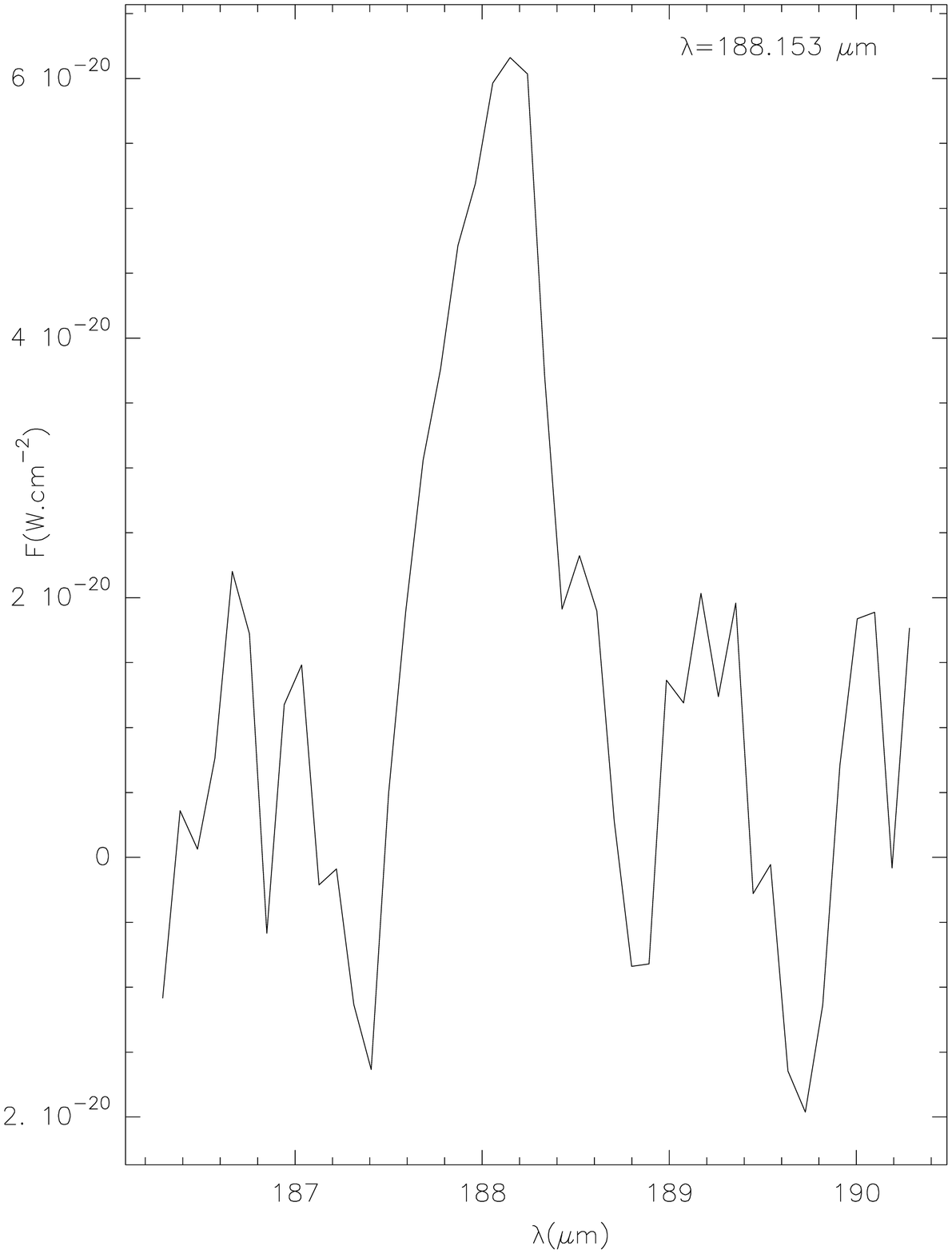} 
\smallskip 
{\small\noindent Fig. 4 --The HCN J=18$-$17 line emission detected by the LWS
grating in CRL 2688 at a resolution of 0.6 $\mu$m (S$/ \sim 4$).} 
\label{fig5} 
\smallskip

\subsection{Low-excitation CO lines}

Two outflows can be seen in the CO line profiles of CRL 618 (see Table 2).
The main outflow
is centered at $-$22.0 \kms $\;$ with an expansion velocity
of roughly 20 \kms.
The optically thick CO J=1$-$0 line clearly absorbs the continuum
from $-$60 to $-$40 \kms $\;$
coinciding with the main outflow terminal velocity. Other lines (higher
transitions) show an absorption dip (self-absorption). In CRL 618, many
molecular lines show P-Cygni profiles at 3 and 2 mm (Cernicharo \etal 2002).
These lines, which
correspond to very high energy ro-vibrational levels of HC$_3$N and
HC$_5$N are formed in the PDR of the high density torus surrounding the
central object.

The {\em HVW} (around 200 \kms), associated with a biconical outflow, is also
observed in many other  molecular transitions (see also
Cernicharo \etal 1989, Cernicharo et al. 2002) but with a smaller 
velocity span.
The CO wings corresponding to this outflow have a smooth triangular
shape and are present in all the millimeter and submillimeter lines. Their
intensity increases with J, $R_{I}$ going from 0.09 for J=1$-$0 to an
estimated value of more than 2.6 in the J=7$-$6 line. Therefore, the
emission in the high J lines (next paragraph) is basically
dominated by this high velocity wind component.

\begin{table*} [ht]
   \caption{ \label{tablex2}Table of the parameters for the different flows in
     CRL 618. The uncertainty on $v_{main}$
     and $v_{wing}$ is 0.3 \kms.
     $R_{I}$ is the wing and main integrated intensities ratio.
     For the 6$-$5 and 7$-$6
lines, the parameters $v_{Wing}$ and $R_{i}$ are lower limits due to the
unsufficient bandwidth available for these observations. }
{\small {\begin{tabular}{l|c|c|c|c|c}
  & \multicolumn{5}{c}{\bf{CRL 618}} \\
  {\emph CO parameters} & 1$-$0 & 2$-$1 & 3$-$2 & 6$-$5 & 7$-$6 \\ \tableline
$v_{main}$ (\kms) & 19.7 & 19.6 & 18.9 & 17.1 & 18.8 \\
$v_{wing}$ (\kms) & 202.5 & 204.5 & 178.5 & $>$76.0 & $\geq195.0$ \\
$ R_{I}$ & 0.09 & 0.22 & 0.45 & $>$0.72 & $\geq$2.6 \\
  $\Delta$$R_{I}$ & (0.01) & (0.02) & (0.03) & (0.05) & (0.3) \\ \tableline
\end{tabular}}}
\end{table*}

\subsection{Far-IR emission}

The rotational lines of \COd (J=14$-$13 to J=41$-$40),
\COt (J=14$-$13 to J=19$-$18), HCN, HNC, [O$\small I$] (at 63.170 and
145.526 $\mu$m) and [C$\small{II}$] ($^{2}p_{3/2}- ^{2}p_{1/2}$ transition at
157.74 $\mu$m) are observed in CRL 618 (see Fig. 3).
Several lines of OH and \water are also detected.
However, no H$_2$O emission has been detected at 183 GHz.
As the 22, 183 and 321 GHz
water maser lines are collisionally excited (radiative effects are small)
in warm dense gas (Neufeld \& Melnick 1990, 1991), this non-detection
may indicate that water is not in the wind region, but rather in the torus.

The model used for analyzing the LWS observations of CRL 618 was
presented in Herpin \& Cernicharo (2000). Briefly, it consists of
a central high density torus with its inner part being a
Photo-Dissociated Region, an extended AGB remnant envelope and high
velocity bipolar outflows. The physical parameters of the different
layers are given in Table 5. The AGB remnant parameters are derived from
Yamamura \etal (1994) and Hajian \etal (1996). The HVW 
characteristics are from Cernicharo \etal (1989), Neri \etal (1992) 
and Martin-Pintado \etal (1995). Justtanont \etal (2000) fit the ISO 
CO lines with a 700 K component
(n(H$_2$)=5 $10^7$ cm$^{-3}$).
Temperatures of the torus layers have been varied in order to get 
the best fit of the data.
Our model, once applied to the low-excitation CO lines,
disagrees with the observed J=3$-$2 to 7$-$6 intensities. This may be
mainly explained by the completely non-spherical geometry of the
source, and by the complex velocity field across the envelope.
Moreover, as the CO molecules are reprocessed, excitation is more complex
now and cannot be reproduced by our relatively simple model.

\subsection{Changes since the previous stage}

As in the previous very young PPN phase, shock chemistry plays an
important role, as even faster outflows are detected. The now
higher effective temperature of the central star induces a UV-based
chemistry dominated by the photolysis of key molecules and
the molecular content has started to be reprocessed. This fact
is proved by the observed changes in molecular abundances
(cf.  HCN, HNC), the huge abundance of cyanopolyynes, the presence
of polyynes and benzene (Cernicharo et al., 2001a\&b) and
of O-bearing molecules (Herpin and Cernicharo 2000). Ionic ([C$\small{II}$])
and atomic fine structure lines ([O$\small{I}$])
are also detected as a result of the photodissociation of CO.
At this stage, the main cooling of the gas occurs via
CO high-J, [C$\small{II}$] and [O$\small{I}$] lines.

\section{A young PN object: NGC 7027}

NGC 7027 is the most evolved object in this study.
It is a very young PN, having left the AGB only $10^{3}$
years ago (Volk \& Kwok 1997). The estimated temperature
of the central star is around 200 000 K (Latter \etal 2000).
The gas in the inner 10$^{\prime\prime}$ radius
of the envelope has been completely ionized and a strong continuum
emission arises from it. This region is revealed in the far-infrared
by fine-structure lines (Liu \etal 1996). The PDR at the interface
between the ionized and neutral regions is mainly traced by the pure
rotational lines of CH$^+$ (Cernicharo \etal 1997). The
H$_2$ emission at this interface is found to be consistent with
excitation by UV photons (Cox \etal 1997).

\subsection{Low-excitation CO lines}

All the CO lines are asymmetrical (see Fig.1), reflecting the complex
kinematical structure of this PN.
The main CO line emission is centered
at $V_{LSR} \sim 27-30$ \kms, with an expansion velocity of 15 \kms 
$\;$ for the
J=1$-$0, 2$-$1, 3$-$2 and 4$-$3 lines, and 20 \kms $\;$ for the
6$-$5 and 7$-$6 lines (see Table 3). The higher velocity extent of the last two
transitions probably indicates the presence of a faster wind at smaller radii
(3$^{\prime\prime}-5^{\prime\prime}$). Jaminet \etal (1991) found 
$v_{exp}\sim 15$ \kms $\;$
(with an angular extension of 5$^{\prime\prime}-15^{\prime\prime}$) 
for the 2$-$1 and 3$-$2 lines in
agreement with our data for those transitions. They also have derived a
velocity dispersion as high as 3 \kms $\;$ at small radii 
($<$5$^{\prime\prime}$),
and more than 1 \kms $\;$ in the outer parts ($>$5$^{\prime\prime}$).
An absorption feature at $v_{LSR} \sim 10$ \kms,
due to self-absorption (Jaminet \etal 1991),
is detected in the low-J lines (see Fig. 1).

\begin{table*} [ht]
   \caption{ \label{tablex3} Table of the parameters for the different flows in
     NGC 7027. The uncertainty on $v_{main}$
     and $v_{wing}$ is 0.3 \kms.
     $R_{I}$ is the wing and main integrated intensities ratio.}
{\small {\begin{tabular}{l|c|c|c|c|c|c}
  & \multicolumn{6}{c}{\bf{NGC 7027}} \\
  {\emph CO parameters} & 1$-$0 & 2$-$1 & 3$-$2 & 4$-$3 & 6$-$5 & 
7$-$6 \\ \tableline
$v_{main}$ (\kms) & 14.6 & 14.3 & 16.0 & 18.3 & 20.0 & 20.0  \\
$v_{wing}$ (\kms) & 47.2 & 43.2 & 30.0 & 25.1 & 24.7 & 23.8 \\
$ R_{I}$ & 0.04 & 0.09 & 0.07 & 0.07 & 0.06 & 0.09  \\
  $\Delta$$R_{I}$ & (0.01) & (0.01) & (0.02) & (0.02) & (0.02) & 
(0.02) \\ \tableline
\end{tabular}}}
\end{table*}

Although the J=1$-$0 line of CO is certainly slightly contaminated
by the 38$\alpha$ recombination line of atomic
hydrogen (shifted only by $-$8.3 \kms $\;$ with respect to its center), the
line profiles show also some
emission at high velocity (mainly in the red part)
that could be the signature of a faster outflow,
with an estimated velocity extent of 45 \kms. This
outflow component was not reported by Jaminet \etal (1991). They only
found the extended outflow with $v_{exp} \sim 15$ \kms $\;$
and a wind at a smaller radius with $v_{exp} \sim$ 23 \kms, which
may be a flow driven by the pressure of the ionized
region, and which is probably the place where the 6$-$5 and
7$-$6 lines are mainly excited (see $v_{main}$ in Table 3). The flow at
higher velocity ($\sim$47 \kms) may be the relic of a {\em High Velocity
Wind} ejected during the PPN phase.
Nevertheless, the importance of the wing
emission is quite low as the ratio $R_{I}$ is less than 0.1.

\subsection{Far-IR emission}

The J=14$-$13 to J=23$-$22 \COd lines are detected in the
LWS grating spectrum of NGC 7027 and some \COt
lines are also marginally detected.
Despite its carbon rich chemistry,
several lines of OH are clearly identified (Liu et al. 1997),
while no H$_2$O emission is observed (Cernicharo et al. 1997).
In addition to the molecular emission, strong  atomic and ionic features
typical of a PN spectrum are
also seen. They include [O$\small I$], [C$\small{II}$], [N$\small{II}$],
[N$\small{III}$] and [O$\small{III}$] lines. The pure rotational lines
of CH$^{+}$  discovered by Cernicharo et al. (1997)
are prominent through the far-infrared spectrum.
Features at 149.18 and 180.7 $\mu$m correspond to rotational lines of
CH (see Liu \etal 1997). A new feature of this species at 115.9 $\mu$m
is tentatively detected.
However, no HCN or HNC emission is seen, which supports
the small HCN abundance ($9 ~10^{-9}$) derived by Deguchi \etal
(1990). The main cooling agents are fine-structure atomic and
ionic lines.

To fit the far-infrared spectrum of NGC7027 we have considered
3 components (see Table 6):
(i) the atomic region; (ii) a high density layer; (iii) the AGB remnant.
Sizes of the different regions are derived from Yan \etal (1999) 
and Jaminet \etal (1991).

The atomic region (i) has been subdivided into two layers of 1000 K and 800 K
and $\theta$= 5$^{\prime\prime}$ and 6$^{\prime\prime}$ respectively
(Cernicharo \etal 1997, Liu \etal 1996, Hasegawa, Volk \& Kwok 2000).
The density is 5 $10^{7}$ cm$^{-3}$, and the expansion velocity 15 \kms.
In the hottest layer, only [O$\small I$] and CH$^{+}$ are present (for now
we have not considered [C$\small{II}$] and
[O$\small{III}$], in the model although they are also present).
In the external layer, we introduce \COd, \COt and OH.
The considerable abundance of OH suggests
that H$_2$O was probably present in the previous stage of
evolution (see CRL 618), and, even if not detected,
may be still present here,
but with a low abundance. Thus we placed H$_2$O in
region (i) with an upper limit of abundance relative to H$_2$ of
  $1.5 ~10^{-7}$, according to the present data. This upper limit is 50\%
lower than the one obtained by Volk \& Kwok (1997). The best
fit to the data gives the following abundances (relative to \COd) in this
layer: $<1/40$, 1$/$20,
1$/$80 and 875 for \COt, OH, CH$^{+}$ and [O$\small I$] respectively.
Comparing with analyses performed by other authors we point out
that the \Htwo
densities used in our model are higher than those of
Liu \etal (1996), who took values between $10^{5}$ and $10^{6}$ cm$^{-3}$ for
the high-J CO lines coming from the hot region (1000$-$700 K). Also, the
velocity of the expanding ionized region derived by Jaminet \etal (1991)
is 17.6 \kms $\;$, while the estimated CO velocity wind by Sopka \etal (1989)
is 17 \kms. Note that the two components of the atomic region 
produce similar lines, except that the hottest one only contains 
[O$\small I$]
and CH$^+$.

In the high density layer (ii) we consider a temperature
of 270 K, $\theta$= 9$^{\prime\prime}$,
expansion velocity of 15 \kms $\;$ and a density of $10^{6}$ cm$^{-3}$
(Yan \etal 1999, Justtanont \etal 2000).
Volk \& Kwok (1997) applied for this layer very similar values
($n_{H_{2}}=9 ~10^{5}$ cm$^{-3}$ and $T=220$ K).
The fit results in a \COd column density of $2.1 ~10^{17}$ cm$^{-2}$
and [\COd$/$\COt]$\leq$30.

The AGB remnant region (iii) has an extension going from  10$^{\prime\prime}$
  to 20$^{\prime\prime}$,
the gas temperature decreases from 100 K to 50 K, and the density from
$3 ~10^{5}$ to $10^{4}$ cm$^{-3}$ (Hasegawa, Volk \& Kwok 2000).
The expansion velocity is 15 \kms $\;$
(see Sect. 6.1 and Volk \& Kwok 1997, Jaminet \etal 1991). The colder layer
(50 K) is important here only to explain the low-J CO emission. The \COd column
density is 1.5 to $1.7 ~10^{17}$ cm$^{-2}$ and
the [\COd$/$\COt] ratio is always $<30$. Note that CO emission
extends over a region larger than 40''(Masson \etal 1985).

Our lower limit of the $^{12}$C/$^{13}$C ratio (30) is compatible with
the value derived by Kahane \etal (1992) for this object (65) using
much more sensitive millimeter observations of several carbon-bearing
molecules. The whole CO column density derived by Thronson
(1983) is larger than $2 ~10^{17}$ cm$^{-2}$, in good agreement with our
values.

The results of our model when applied to the low-excitation CO lines are
satisfactory for J=1$-$0 to J=4$-$3, but the predicted intensities are too high
for the J=6$-$5 and 7$-$6 lines. The reason might be the complex 
kinematical structure of this PN.

\subsection{Changes since the previous stage}

The main characteristic of this young PN compared with previous
stages is the almost disappearance of fast molecular winds.
The hot central object induces a strong UV field leading to a UV-dominated
photochemistry in the envelope. As a consequence,
molecular abundances have been strongly
modified since the PPN phase. Water vapor and HCN have almost disappeared,
while other species such as CH$^{+}$ have been efficiently formed.
The cooling effect of CO lines is much weaker and now the
gas is mainly cooling via ionic and atomic fine structure lines.

\begin{table*} [h]
   \caption{ \label{table1}Table of the column densities (in 
cm$^{-2}$) obtained from our model for all
molecules present in each region of CRL 2688
($\theta$ defines the solid angle of each considered layer).}
   {\small{\begin{tabular}{l|c|c|c} \hline \hline
  {\bf Molecules} & {\bf \COd} & {\bf \COt} & {\bf HCN} \\ \tableline
{\bf FAST WIND FROM THE STAR (i)} & & & \\
T=800 K, $\theta= 0.3''$, $v=200$ \kms, $n(H_{2})=$5 $10^{7}$ cm$^{-3}$
& 2 $10^{18} $ & $<$8 $10^{16}$ & $<$2 $10^{16}$ \\
T=500 K, $\theta= 2''$, $v=100$ \kms, $n(H_{2})=$5 $10^{7}$ cm$^{-3}$
& 7 $10^{17} $ & $<$2.8 $10^{16}$ & $<$2.3 $10^{16}$ \\
T=400 K, $\theta= 3''$, $v=60$ \kms, $n(H_{2})=$5 $10^{7}$ cm$^{-3}$
& 5 $10^{17} $ & $<$2 $10^{16}$ & $<$1.7 $10^{16}$  \\ \hline
{\bf MEDIUM WIND FROM THE STAR (ii)} & & & \\
T=250 K, $\theta= 15''$, $v=40$ \kms, $n(H_{2})=$5 $10^{7}$ cm$^{-3}$
& 4.5 $10^{16} $ & $<$1.8 $10^{15}$ & $<$1.5 $10^{15}$  \\ \hline
{\bf AGB remnant (iii)} & & & \\
T=50 K, $\theta= 25''$, $v=20$ \kms, $n(H_{2})=$5 $10^{5}$ cm$^{-3}$
& 3.5 $10^{16} $ & $<$1.4 $10^{15}$ & $<$1.2 $10^{15}$  \\
\tableline
  \end{tabular}}}
\end{table*}
%


\begin{table*} [h]
   \caption{ \label{table2}Table of the column densities (in cm$^{-2}$)
obtained from our model for all
molecules present in each region of CRL 618
($\theta$ defines the solid angle of each considered layer).}
   {\small{\begin{tabular}{l|c|c|c|c|c|c|c} \tableline \tableline
  {\bf Molecules} & {\bf \COd} & {\bf \COt} & {\bf HCN} & {\bf HNC} & {\bf
o-\water}
& {\bf OH} & {\bf [O$\small I$]} \\ \tableline
{\bf TORUS} (including PDR) & & & & & & & \\
T=250 K, $\theta= 1.5''$ & $10^{19} $ & 5 $10^{17}$ & $10^{16}$
& $10^{16}$ & 3 $10^{17}$ & 8 $10^{15}$ & \\
$v=20$ \kms, $n(H_{2})=$5 $10^{7}$ cm$^{-3}$ & & & & & & & \\
T=800 K, $\theta= 1.1''$ & 6 $10^{17} $ & 3 $10^{16}$ & & & & & \\
$v=20$ \kms, $n(H_{2})=$5 $10^{7}$ cm$^{-3}$ & & & & & & & \\
T=1000 K, $\theta= 0.6''$ & $10^{19} $ & 5 $10^{17}$ & & & &
& 4.5 $10^{19}$  \\
$v=20$ \kms, $n(H_{2})=$7 $10^{7}$ cm$^{-3}$ & & & & & & & \\  \hline
{\bf HVW} & & & & & & & \\
T=200 K, $\theta= 1.7''$ & 5 $10^{18} $ & 2.5 $10^{17}$ & 3 $10^{17}$
& 2 $10^{17}$ & & & \\
$v=50$ \kms, $n(H_{2})= 10^{7}$ cm$^{-3}$ & & & & & & & \\
T=1000 K, $\theta= 1.5''$ & 2 $10^{16} $ & 1 $10^{15}$ &  2 $10^{15}$
& $10^{14}$ & & & \\
$v=200$ \kms, $n(H_{2})=10^{7}$ cm$^{-3}$ & & & & & & & \\  \hline
{\bf AGB remnant} & & & & & & & \\
T=50 K, $\theta= 10''$ & 7 $10^{17} $ & 3.5 $10^{16}$ & 7 $10^{16}$
& 7 $10^{14}$ & & & \\
$v=20$ \kms, $n(H_{2})=$ $10^{5}$ cm$^{-3}$ & & & & & & & \\
\tableline
  \end{tabular}}}
\end{table*}
%


\begin{table*} [h]
   \caption{ \label{table3}Table of the column densities (in cm$^{-2}$)
obtained from our model for all
molecules present in each region of NGC 7027
($\theta$ defines the solid angle of each considered layer).}
   {\small{\begin{tabular}{l|c|c|c|c|c|c} \tableline \tableline
  {\bf Molecules} & {\bf \COd} & {\bf \COt} & {\bf o-H$_{2}$O} & {\bf OH}
& {\bf [O$\small I$]} & {\bf CH$^{+}$} \\ \tableline
{\bf ATOMIC REGION (i)} & & & & & & \\
T=1000 K, $\theta= 5''$ & & & & & 2 $10^{19}$ & 9 $10^{13}$ \\
$v=15$ \kms, $n(H_{2})=$5 $10^{7}$ cm$^{-3}$
& & & & & & \\
T=800 K, $\theta= 6''$
& 8 $10^{15} $ & 2.7 $10^{14}$ & $<$2.6 $10^{13}$ & 4 $10^{14}$ & 7 $10^{18}$ &
$10^{14}$ \\
$v=15$ \kms, $n(H_{2})=$5 $10^{7}$ cm$^{-3}$
& & & & & & \\ \hline
{\bf HIGH DENSITY LAYER(ii)} & & & & & & \\
T=270 K, $\theta= 9''$ & 2.1 $10^{17} $ & 7 $10^{15}$ & & & & \\
$v=10$ \kms, $n(H_{2})=10^{6}$ cm$^{-3}$ & & & & & & \\ \hline
{\bf WIND REGION (iii)} & & & & & & \\
T=100 K, $\theta= 10''$ & 1.5 $10^{17} $ & 5 $10^{15}$ & & & & \\
$v=15$ \kms, $n(H_{2})=$3 $10^{5}$ cm$^{-3}$ & & & & & & \\
T=50 K, $\theta= 20''$ & 1.7 $10^{17} $ & 5.7 $10^{15}$ & & & & \\
$v=15$ \kms, $n(H_{2})=10^{4}$ cm$^{-3}$ & & & & & & \\
\tableline
  \end{tabular}}}
\end{table*}


\begin{table*} [h]
   \caption{ \label{tablex} Observed and predicted intensities for the 
mm and submm CO lines for each source.}
{\small {\begin{tabular}{lr|c|c|c|c|c|c}
  {\bf {Source}} & & \multicolumn{6}{c}{\bf{CO lines}} \\ \tableline
   & & 1$-$0 & 2$-$1 & 3$-$2 & 4$-$3 & 6$-$5 & 7$-$6  \\ \tableline
  CRL 2688 & I$_{obs}$(Jy) & 30 & 70 & 271 & 373 & 640 & 760  \\
  & I$_{mod}$(Jy) & 10 & 30 & 296 & 429 & 650 & 790 \\ \tableline
  CRL 618 & I$_{obs}$(Jy) & 23 & 47 & 79 &  & 320 & 232  \\
  & I$_{mod}$(Jy) & 21 & 66 & 197 & 400 & 742 & 994 \\ \tableline
  NGC7027 & I$_{obs}$(Jy) & 36 & 70 & 364 & 516 & 520 & 840  \\
  & I$_{mod}$(Jy) & 24 & 83 & 475 & 498 & 893 & 1061 \\ \tableline
\end{tabular}}}
\end{table*}

\section{Discussion}

$\bullet$ It is assumed that CRL 2688 has just left the AGB phase:
the central object is not yet sufficiently hot to produce a
significant ionization or photodissociation of the gas.
Hence, the far-IR molecular emission
(\COd, \COt, HCN) is  well reproduced by our model.
Nevertheless, the detection of the polyynes C$_4$H$_2$ and C$_6$H$_2$ in
this object indicates that some chemical processing has already
started, an important difference as respects to IRC+10216 where
these species are not detected.

In addition to the low expanding AGB CSE remnant, this PPN exhibits an
inner and faster outflow which runs into that AGB envelope. This
is one of the main characteristics of an object beginning its transition
to the PN stage. At this point, emission from the molecular gas is the
main cooling process.

The \COt emission is not detectable at the low sensitivity of the LWS grating
spectrum. For a \COd abundance of 6 ~$10^{-4}$ the upper limit of
\COt abundance would be $<2.2~ 10^{-5}$.
This value is half of what was found by Yamamura \etal 1996 ($4 ~10^{-5}-7.5
~10^{-5}$), but is similar to the \COt abundance found in an AGB star
such as IRC+10216 (1-3 $10^{-5}$, Cernicharo \etal 1996). However, Kahane
\etal (1992) derived more accurate isotopic ratios, $3.7 ~10^{-5}$ and
$7.5 ~10^{-5}$, respectively for IRC+10216 and CRL 2688, showing
an increase by a factor of 2.
For Jaminet \etal (1992), CNO nuclear processing in the star injects material
into the stellar wind: in the slow wind, $[\COd/\COt]=20$, while in the fast
wind they find a ratio of 5. Because of these values, these authors
placed CRL 2688 among the J-type
carbon stars, which have a considerably lower $\COd/\COt$ ratio than
carbon stars (40$-$80). However, we prefer to argue that CRL 2688 has just
begun to evolve to the PN stage; the \COd material has just began to be
reprocessed, as proved by the apparent increase of the \COt
abundance between IRC+10216 and CRL 2688.

The HCN$/$CO ratio in IRC+10216 is 0.1,
much larger than in the high velocity winds of the post-AGB envelopes
discussed here. According to our model, we find a lower HCN abundance
(relative to \COd) in the fast wind ($\leq 6 ~10^{-6}$) than in the slow
wind ($\leq 2 ~10^{-5}$, i.e., similar to IRC+10216).
Sopka \etal (1989) found $3.3 ~10^{-6}$ in the fast wind
while Jaminet \etal (1992) found $7 ~10^{-7}$.

HNC is not detected in this object. This molecule is mainly
formed by ion-molecule reactions, and in this post-AGB envelope, physical
conditions are therefore not very suitable for a considerable HNC production.
Moreover, Yamamura \etal (1996) and Sopka \etal (1989) derived [HCN$/$HNC]
$\sim 21$ and 160 respectively. According to this fact, the HNC emission
is largely under our detection limit.

$\bullet$ In CRL 618, Herpin \& Cernicharo (2000) have shown
that O-bearing species, \water and OH, are produced in the innermost
region of the circumstellar envelope. Also Cernicharo \etal (2001 a\&b)
have detected in this object the poli-acetylenic
chains C$_{4}$H$_{2}$ and C$_{6}$H$_{2}$, methyl-polyynes, and benzene.
The UV photons from the central star photodissociate most of the
molecular species produced in the AGB phase and allow a chemistry dominated
by standard ion-neutral reactions.
O-bearing species are formed, and also
abundances of C-rich molecules such as HCN and HNC are modified
(see Table 7). At the high
temperatures of the PDR, OH can be formed by endothermic reactions
between O and \Htwo, with O released by CO photodissociation, and
then can produce \water through
reactions with \Htwo. \water and OH can also be produced
from the dissociative recombination of H$_3$O$^+$ if this ion is
abundant enough.

$\bullet$ NGC 7027 has a very hot central star.
Because of the high UV
flux ($10^{4} \leq G_{0} \leq 10^{5}$, Burton \etal 1990),
there is a strong presence of atomic lines. UV photons from the central star
produce a PDR and the gas cools mainly via fine-structure atomic lines.
In the PDR, the total number of CO molecules is around 2$\%$ of that
of ionized carbon, i.e. most of the CO molecules
have been photodissociated (Liu \etal 1996). As the ionized region is
not modeled in this work, our results concerning the atomic region have to be
carefully taken: in the high electron density environment of NGC 7027, ionic
fine-structure lines are strongly suppressed by collisional de-excitation
(Liu \etal 1996, Keyes \etal 1990). Electron impact may be important in
exciting the CH$^{+}$ lines and would lower the CH$^{+}$ abundance (Cernicharo
\etal 1997). In the 1000 K layer, the formation of CH$^{+}$ may be achieved
via $C^{+} + H_{2} \rightarrow CH^{+} + H $ (activation energy
of $\sim$ 4000 K).
The formation rate of CH$^{+}$ (Sternberg \& Dalgarno 1995)
could be high enough to
produce column densities similar to those derived in this work.
Due to the high temperatures,
densities and UV radiation field in the PDR, the
formation of CH$^{+}$ leads to the creation of CH$_{2}^{+}$ and CH$_{3}^{+}$,
whose dissociative recombination will form CH very efficiently at a
temperature of 800 K (Sternberg \& Dalgarno 1995); Liu \etal (1997) derived
[CH$/$CH$^{+}$]$\sim 0.2$. The observed CH lines are excited by collisions
with atomic and molecular hydrogen.

The presence of [O$\small I$] and CH$^{+}$ in the
1000 K layer of the atomic region, and the low abundance of CO in the
800 K layer, indicate that the CO molecules have been largely reprocessed
there through UV photons. Note
that the chemical model of Hasegawa, Volk \& Kwok (2000) shows that
the CO observed in these objects consists of new formed molecules. 
Nevertheless,
the AGB circumstellar gas has not been totally reprocessed: the ionized and
photodissociation regions still constitute a relatively small but
significant fraction of the total mass of the circumstellar
material around NGC 7027.

The study of CRL 618 showed that \water and OH can be efficiently produced.
However, these O-bearing species are also quickly reprocessed. This is
seen in the next stage, represented by NGC 7027. The equilibrium models
(McCabe, Connon, Smith \& Clegg 1979) predict essentially no water
in the carbon-rich conditions of the NGC 7027 envelope.
Volk \& Kwok (1997) argue that even under non-equilibrium
chemical conditions, only a very small fraction
of oxygen will be converted to water here.
The non detection of \water and the detection of OH in
NGC 7027 indicate that \water molecules formed
in the previous stage may have been almost entirely reprocessed.
In the NGC 7027 environment, the simplest explanation for the water vapor
disappearance is its transformation into OH and H by UV photo-dissociation.
This will explain the increase of OH abundance by a factor 60 since
the previous PPN stage represented by CRL 618.
Concerning HCN, probably most molecules have been photodissociated
into  H and CN (CN was detected by Bachiller
\etal 1997 with a strong abundance, CN$/$HCN=9).
Photo-dissociation of CO, HCN, and other carbon-bearing molecules
contribute to the increase of carbon ions abundance.

\begin{table*} [h]
   \caption{ \label{table4}Table of the molecular, atomic and ionic
     abundances (relative to \COd)
     in CRL 2688, CRL 618 and NGC 7027 as seen by the model.}
   {\small{\begin{tabular}{c|c|c|c} \tableline \tableline
  {\bf Species} & {\bf CRL 2688} & {\bf CRL 618} & {\bf NGC7027} \\ \tableline
\COt & $<1/25$ & $1/20$ & $\lesssim 1/30$ \\
HCN & $<1/30-1/100$ & $1/10-1/1000$ & \\
HNC & & $1/10-1/1000$ & \\
\water & & $1/25$ & $<1/650$ \\
OH & & $1/1250$ & $1/20$  \\
$[$OI$]$ & & 4.5 & 875 \\
CH$^{+}$ & & & 1/80 \\
\tableline
  \end{tabular}}}
\end{table*}

\section{Conclusion}

The study of the selected 3 objects in rapid evolution from the AGB to the PN
stage clearly shows the crucial importance that (i)
the interaction between fast and slow winds; and (ii)
the increase of UV flux as the central object evolves toward the white dwarf
state have on the chemistry. The spectroscopic evidence of the 
increasing UV flux
comes from the appearance of atomic and ionic
lines in CRL 618, which later dominate the far-IR spectrum of NGC 7027.
Atomic and ionic lines tend to appear (at ISO's sensitivity) when the
central star is hotter than 10000 K (Fong et al. 2001; Castro--Carrizo et
al. 2001).
On the other hand, the spectroscopic signature showing that
shocks become less important as the evolution goes on, comes
from the wind velocity decrease seen in the CO profiles (this must
be confirmed by further studies on more objects and only refers to the
neutral gas). The strongest shocks occur just
after leaving the AGB when the central star is ejecting large
amounts of material in a very fast wind (the case of CRL 2688).
The AGB remnant envelope is being shocked by an inner, faster
wind developed in the PPN stage (i.e., the 200 \kms $\;$ wind in CRL 618).
Most of the \COd, \COt and HCN emission is produced in these shocks.
%
%
The fast increase of the stellar temperature, will produce a new UV-dominated
chemistry when T$_{eff}\geq$30000 K. These new conditions
(UV photons and shocks) will deeply modify the constitution of the inner
parts of the
envelope. Indeed, in CRL 618, O-bearing molecules (\water and OH, Herpin
and Cernicharo 2000) and relatively complex organic
molecules (Cernicharo et al., 2001a\&b) appear. Furthermore,
most of the  CO and HCN will be entirely reprocessed in the PDR 
leading to strong
HNC emission. At this point, CO and [O$\small I$] atomic lines are the
dominant coolants.

As the star reaches the PN stage, the strong fast molecular
winds have disappeared, and slow expanding layers constitute the PN envelope
around a large and hot atomic region. Most of the {\em old} AGB material
has been reprocessed.
The spectrum is now dominated by atomic and ionic lines. New species
such as CH$^{+}$ and CH appear. There is only weak HCN emission, as the
molecules may have been broken into H and CN. More interesting
is the disappearance of \water, which has probably also
been reprocessed, and is only a relatively
abundant molecule in the intermediate C-rich PPN stage. \\

{\it Acknowledgments}
We thank Spanish DGES, CICYT, and PNIE for funding support
for this research under grants PB96-0883, ESP98-1351E and
PANAYA2000-1784.
JRG acknowledges \textit{UAM} for a
pre-doctoral fellowship. JRP also acknowledges further support
for his research from NSF grants AST99-80846 (CSO operations)
and ATM96-16766. We thank Dr. V. Bujarrabal for useful comments
and suggestions. We thank the anonymous referee for his useful comments which 
result in an important improvement of this paper.


\begin{references}


\reference{}
Acker, A. \etal 1992, Strasbourg-ESO catalog of galactic planetary nebulae,
ESO, Garching

\reference{}
Bachiller, R., Forveille, T., Huggins, P.J., Cox, P. 1997, A\&A 324, 1123

\reference{}
Bujarrabal, V., G\'{o}mez-Gonz\'{a}lez, J., Bachiller, R., Mart\'{\i}n-Pintado,
 J. 1988, A\&A 204, 242

\reference{}
Bujarrabal, V., \& Alcolea, J. 1991, A\&A, 251, 536

\reference{}
Bujarrabal, V., Fuente, A., Omont, A. 1994, A\&A 285, 247

\reference{}
Bujarrabal, V., Castro--Carrizo, A., Alcolea, J., \& S\'anchez--Contreras, C.
2001, A\&A, 377, 868

\reference{}
Burton, M.G., Hollenbach, D.J., Tielens, A.G.G.M. 1990, ApJ 365, 620

\reference{}
Carsenty, U., Solf, J. 1982, A\&A 106, 307

\reference{}
Castro-Carrizo, A., Bujarrabal, V., Fong, D. \etal 2001, A\&A 367, 674


\reference{}
Cernicharo, J., Gu\'{e}lin, M., Mart\'{\i}n-Pintado, J., Pe$\tilde{n}$alver, J.,
Mauersberger, R. 1989, A\&A 222, L1

\reference{}
Cernicharo, J., Barlow, M.J., Gonz\'alez-Alfonso, E., \etal 1996, A\&A 315,
L201

\reference{}
Cernicharo, J., Liu, X.W., Gonz\'{a}´lez-Alfonso, E., Cox, P., Barlow, M.J.,
Lim, T., Swinyard, B.M. 1997, ApJ 483, L65

\reference{}
Cernicharo, J., Yamamura, I., Gonz\'alez-Alfonso, E., \etal 1999, ApJ 526, L41

\reference{}
Cernicharo, J., Gu\'elin, M., Kahane, C. 2000, AASS 142, 181

\reference{}
Cernicharo, J., Heras, A.M., Tielens, A.G.G.M., \etal 2001a, ApJ Letters
546, L123

\reference{}
Cernicharo, J., Heras, A.M., Pardo, J.R., \etal 2001b, ApJ Letters 546, L127

\reference{}
Cox, P., Gonz\'{a}lez-Alfonso, E., Barlow, M.J., \etal 1996, A\&A 315, L265

\reference{}
Cox, P., Maillard, J.P., Huggins, P.J., \etal 1997, A\&A 321, 907

\reference{}
Cox, P., Lucas, R., Huggins, P.J., Forveille, T., Bachiller, R., Guilloteau, S.,
Maillard, J.P., Omont, A., 2000 A\&A 353, L25

\reference{}
Crosas, M., Menten, K.M., 1997 ApJ 483, 913

\reference{}
Deguchi, S., Izumiura, H., Kaifu, N., Mao, X., Nguyen-Q-Rieu, Ukita, N., 1990
ApJ 351, 522


\reference{}
Fong, D., Meixner, M., Castro-Carrizo, A., Bujarrabal, V.,
Latter, W.B., Tielens, A.G.G.M., Kelly, D.M., Sutton, E.C. 2001, A\&A, 367, 652

\reference{}
Frank, A., Balick, B., Icke, V., Mellema, G., 1993 ApJ 404, L25

\reference{}
Fuente, A., Cernicharo, J., Omont, A., 1998 A\&A 330, 232


\reference{}
Glassgold, A.E. 1996, ARAA 34, 241


\reference{}
Groenewegen, M.A.T. 1997, A\&A 317, 503

\reference{}
Groenewegen, M.A.T., van der Veen, W.E.C.J., Mattheus, H.E. 1998, A\&A 338, 491

\reference{}
Hajian, A. R., Phillips, J. A., Terzian, Y. 1996, ApJ 467, 341

\reference{}
Hasegawa, T., Volk, K., Kwok, S. 2000, ApJ 532, 994

\reference{}
Herpin, F., Cernicharo, J., 2000 ApJ 530, L129

\reference{}
Hora, J.L., Deutsch, L.K., Hoffmann, W.F., Fazio, G.G. 1996, AJ 112, 2064

\reference{}
Jaminet, P.A., Danchi, W.C., Sutton, E.C., \etal 1991, ApJ 380, 461

\reference{}
Jaminet, P.A., Danchi, W.C., Sandell, G., Sutton, E.C. 1992, ApJ 400, 535

\reference{}
Jura, M., Kroto, H. 1990, ApJ 351,222

\reference{}
Justtanont, K., Tielens, A.G.G.M., Skinner, C.J., Haas, M.R. 1997, ApJ 476. 319

\reference{}
Justtanont, K., Barlow, M.J., Tielens, A.G.G.M., \etal 2000, A\&A 360, 1117

\reference{}
Kahane, C., Cernicharo, J., G\'{o}mez-Gonz\'{a}lez, J.,
Gu\'{e}lin, M. 1992, A\&A 256, 235

\reference{}
Kahane, C., Dufour, E., Busso, M. \etal 2000, A\&A 357, 669

\reference{}
Keyes, C.D., Aller, L.H., Feibelman, W.A. 1990, PASP 102, 59

\reference{}
Kooi, J.W., Kawamura, J., Chen, J., \etal 2000, Int. J. of Infrared
and Millimeter Waves, 21, 9

\reference{}
Kwok, S. 2000, in {\em The origin and evolution of planetary nebulae},
Cambridge astrophysics series 31, Cambridge university press

\reference{}
Kwok, S., Su, K.Y.L., Stoesz, J.A. 2001, in {\em Post-AGB objects as a
phase of stellar evolution}, edited by R. Szczerba and S.K. Gorny, Astrophys.
and Space Science Library, vol. 265


\reference{}
Latter, W.B., Dayal, A., Bieging, J.H., \etal 2000, ApJ 539, 783

\reference{}
Liu, X.W., Barlow, M.J., Nguyen-Q-Rieu, \etal 1996, A\&A 315, L257

\reference{}
Liu, X.W., Barlow, M.J., Dalgarno, A., \etal 1997, MNRAS 290, L71

\reference{}
Liu, X.W. \etal 2001, MNRAS 323, 343

\reference{}
Loup, C., Forveille, T., Omont, A., Paul, J.F. 1993, AASS 99, 291

\reference{}
Lucas, R., Gu\'elin, M. 1999, in {\em Asymptotic giant branch stars}, p.305,
IAU symposium 191, published by the Astronomical Society of the Pacific

\reference{}
McCabe, E.M., Connon Smith, R., Clegg, R.E.S. 1979, Nature 281, 263

\reference{}
Martin-Pintado, J., Gaume, R. A., Johnston, K. J., Bachiller, R. 1995, 
ApJ 446, 687

\reference{}
Masson, C.R., Cheung, K.W., Berge, G.L., \etal 1985, ApJ 292, 464

\reference{}
Mauron, N., Huggins, P.J. 1999, A\&A 349, 203

\reference{}
Men'shchikov, A.B., Bolega, Y., Bl\"ocker, T. \etal 2001, A\&A 368, 497

\reference{}
Meixner, M., Fong, D., Sutton, E.C., Welch, W.J. 2001, in {\em Post-AGB 
objects as a
phase of stellar evolution}, edited by R. Szczerba and S.K. Gorny, Astrophys.
and Space Science Library, vol. 265

\reference{}
Neri, R., Garcia-Burillo, S., Gu\'{e}lin, M., Cernicharo, J., Guilloteau, S.,
Lucas, R. 1992, A\&A 262, 544

\reference{}
Neufeld, D., Melnick, G. 1990, ApJ 352 L9

\reference{}
Omont, A., Moseley, S.H., Cox, P. 1995, ApJ 454, 819

\reference{}
Penzias, A.A., Burrus, C.A. 1973, ARA\&A 11, 51

\reference{}
Sahai, R., Trauger, J.T., Watson, A.M., \etal 1998, ApJ 493, 301

\reference{}
Salas, J.B., Pottasch, S.R., Beintama, D.A., Wesselius, P.R., 2001 A\&A 367, 949

\reference{}
Skinner, C.J., Meixner, M., Bobrowsky, M. 1998, MNRAS 300, L29

\reference{}
Sopka, R.J., Olofsson, H., Johansson, L.E.B., Nguyen Q-Rieu, Zuckerman, B. 1989,
A\&A 210, 78

\reference{}
Speck, A.K., Meixner, M., Knapp, G.R. 2000, ApJ 545, L145

\reference{}
Sternberg, A., Dalgarno, A. 1995, ApJS 99, 565

\reference{}
Thronson, H.A. 1983, ApJ 264, 599

\reference{}
Tsuji, T. 1973, A\&A 23, 411

\reference{}
Volk, K., Kwok, S. 1997, ApJ 477, 722

\reference{}
Volk, K., Xiong, G.-Z., Kwok, S. 2000, ApJ 454, 819


\reference{}
Yamamura, I., Shibata, K.M., Kasuga, T., Deguchi, S. 1994, ApJ 427, 406

\reference{}
Yamamura, I., Onaka, T., Kamijo, F., Deguchi, S., Ukita, N. 1996, ApJ 465, 926

\reference{}
Yan, M., Federman, S.R., Dalgarno, A., Bjorkman, J.E. 1999, ApJ 515, 640

\reference{}
Young, K., Serabyn, G., Phillips, T.G. \etal 1992, ApJ 385, 265

\reference{}
Zuckerman, B., Aller, L.H. 1986, ApJ 301, 772

\end{references}
\end{document}